\documentclass[
  reprint,
  amsmath,amssymb,
  aps,
  prapplied
]{revtex4-2}

\usepackage{graphicx}
\usepackage{dcolumn}
\usepackage{bm}
\usepackage{hyperref}
\usepackage[utf8]{inputenc}
\usepackage{upgreek}

\begin{document}

\title{Terahertz-based longitudinal phase space diagnostics of laser wakefield accelerated electron beams}

\author{Zhihao Chen}
\affiliation{Department of Engineering Physics, Tsinghua University, Beijing 100084, China}

\author{Yu Fang}
\affiliation{Department of Engineering Physics, Tsinghua University, Beijing 100084, China}

\author{Ran Li}
\affiliation{Department of Engineering Physics, Tsinghua University, Beijing 100084, China}

\author{Xiancong Yang}
\affiliation{Department of Engineering Physics, Tsinghua University, Beijing 100084, China}

\author{Tianliang Zhang}
\affiliation{Department of Engineering Physics, Tsinghua University, Beijing 100084, China}

\author{Jianfei Hua}
\email{jfhua@tsinghua.edu.cn}
\affiliation{Department of Engineering Physics, Tsinghua University, Beijing 100084, China}

\author{Fei Li}
\affiliation{Institute of High Energy Physics, Chinese Academy of Sciences, Beijing 100049, China}

\author{Wei Lu}
\affiliation{Institute of High Energy Physics, Chinese Academy of Sciences, Beijing 100049, China}
\affiliation{Beijing Academy of Quantum Information Sciences, Beijing 100193, China}

\date{\today}

\begin{abstract}
Femtosecond relativistic electron beams are key probes of ultrafast dynamics, and their pulse duration directly limits the 
achievable temporal resolution. Laser wakefield acceleration (LWFA) provides a compact source of such beams, but the 
injection-induced energy spread makes bunch compression sensitive to nonlinear longitudinal transport, motivating direct 
longitudinal phase space (LPS) measurements. Here, a terahertz transverse-deflecting cavity (THz-TDC) combined with a dipole 
magnet is used to reconstruct the nonlinear LPS of LWFA electron bunches compressed in a double-bend achromat (DBA), 
resolving a characteristic C-shaped distribution associated with higher-order longitudinal transport. At an average energy of 
approximately \(4.55~\mathrm{MeV}\), the diagnostic achieves a temporal resolving power of \(1.8~\mathrm{fs}\) and an energy 
resolution of \(6.0~\mathrm{keV}\), corresponding to a relative energy resolution of \(0.13\%\). For comparable energy 
spreads of approximately \(2.9\%\), shifting the transmitted energy-window center from \(4.574~\mathrm{MeV}\) to 
\(4.532~\mathrm{MeV}\) moves the selected beam away from a low-slope region of the nonlinear LPS and increases the 
root-mean-square bunch length from \(26~\mathrm{fs}\) to \(42~\mathrm{fs}\); with the window center held near 
\(4.553~\mathrm{MeV}\), increasing the energy spread from \(2.0\%\) to \(4.4\%\) lengthens the bunch from \(27~\mathrm{fs}\) 
to \(44~\mathrm{fs}\). These results show that the final bunch duration is governed by both the position and width of the 
transmitted energy window within the nonlinear LPS, establishing an LPS-guided strategy for optimizing DBA-compressed LWFA 
electron bunches and providing a basis for future higher-order phase-space correction.
\end{abstract}

\maketitle

\section{\label{sec:intro}Introduction}

Relativistic electron beams with femtosecond-scale durations are important
probes for resolving transient processes at microscopic scales, with
applications ranging from ultrafast electron diffraction
(UED)~\cite{siwick2003atomic,mo2018heterogeneous,yang2018imaging,yang2020simultaneous}
and microscopy (UEM)~\cite{li2014single,kuwahara2023temporal} to broader
ultrafast electron-based imaging of transient electromagnetic structures, such
as plasma wakefields~\cite{zhang2017femtosecond,wan2024real}. In these
applications, the electron pulse duration is a key factor determining the
achievable temporal resolution, making ultrashort bunch generation and
compression essential. Laser wakefield acceleration (LWFA) has emerged as a
promising compact approach, as the small-scale plasma accelerating structure
intrinsically enables the generation of femtosecond-scale electron
bunches~\cite{khachatryan2007femtosecond,papp2019self,winkler2025active}.
However, the finite energy spread introduced during injection makes these
bunches particularly sensitive to energy-dependent longitudinal transport
during subsequent
propagation~\cite{esarey2009physics,pak2010injection}. In a drift section,
electrons with different energies accumulate different time of flight, leading
to temporal broadening. This longitudinal dispersion is characterized by the
first-order transport coefficient, \(R_{56} = L/\gamma^2\), where \(L\) is the
drift length and \(\gamma\) represents the relativistic Lorentz factor. This
effect is non-negligible for few-\(\mathrm{MeV}\) electron beams, where the
velocity remains sufficiently energy-dependent that, for example, a \(1\%\)
energy spread over a meter-scale propagation distance can increase an initially
few-femtosecond bunch duration to hundreds of femtoseconds.

To mitigate this temporal elongation, beamlines often employ bunch compression
schemes, among which magnetic compression is one of the most widely used
approaches. Magnetic compression utilizes dispersive sections, such as
chicanes~\cite{zhu2016sub,he2015design,lttypes},
\(\alpha\)-magnets~\cite{enge1963achromatic,rajabi2018design,yang2025sub,pires2025comparison},
and double-bend achromats (DBAs)~\cite{fang2022ultrafast,kim2020towards,qi2020breaking},
to introduce energy-dependent path lengths for electrons with different
energies. The resulting longitudinal delay can compensate the velocity-induced
temporal broadening accumulated during beam propagation. In terms of the
first-order longitudinal transport, the compensation condition can be expressed
as
\begin{equation}
R_{56}=R_{56,\mathrm{prop}}+R_{56,\mathrm{comp}}\approx 0,
\label{eq:compensation}
\end{equation}
where \(R_{56,\mathrm{prop}}\) represents the longitudinal dispersion
accumulated during beam propagation, and \(R_{56,\mathrm{comp}}\) denotes the
compensating contribution introduced by the magnetic compression section.
However, as this first-order term is tuned toward zero, second-order effects become increasingly 
important, and the linear compression picture is no longer sufficient to describe the beam 
dynamics. To second order, the longitudinal coordinate can be written as \(z = z_0 + 
R_{56}\delta_0 + T_{566}\delta_0^2\)~\cite{winkler2025active}, where \(z_0\) and \(z\) denote the 
longitudinal coordinates before and after transport, \(\delta_0\) is the initial relative energy 
deviation, and  \(T_{566}\) is the second-order longitudinal transport coefficient. The 
\(T_{566}\) term bends an initially quasi-linear time--energy correlation into a nonlinear 
longitudinal phase space (LPS). Equivalently, the local time--energy slope is given by
\begin{equation}
\frac{\partial z}{\partial \delta_0} = R_{56} + 2 T_{566}\delta_0.
\label{eq:local-slope}
\end{equation}
This relation shows that different energy regions of a nonlinear LPS can have different temporal 
projections, and that a broader energy spread covers a wider range of nonlinear phase space. 
Consequently, the final bunch duration depends not only on the energy spread, but also on the 
position of the energy range relative to a region with a small local time--energy slope.

For LWFA beams, the broad occupied energy range can make the second-order contribution 
\(T_{566}\delta_0^2\) appreciable across the bunch, leading to pronounced nonlinear LPS 
distortions. Directly resolving the LPS is therefore essential not only for identifying the 
nonlinear time–-energy correlation, but also for determining how the accessible energy range 
should be selected to optimize the final bunch duration.

Longitudinal phase-space diagnostics have been widely developed in conventional accelerator 
beamlines to characterize and optimize ultrashort electron bunches. Transverse deflecting 
structures combined with dispersive sections can directly map the temporal and energy 
coordinates of the bunch onto a transverse screen, enabling measurements of the current profile, 
energy chirp, slice energy spread, and two-dimensional 
LPS~\cite{emma2000transverse,marx2018longitudinal}. Tomographic and wakefield-based approaches 
have also been used to reconstruct or infer the 
LPS~\cite{loos2004longitudinal,malyutin2017longitudinal,dijkstal2024longitudinal}.
In parallel, higher-order phase-space manipulation using sextupole magnets, harmonic 
radio-frequency fields, and plasma-based linearizers has been explored to compensate nonlinear 
longitudinal transport or energy 
chirp~\cite{england2005sextupole,vogel2010test,wu2021tunable,wu2023linearization}.
These studies establish the importance of LPS-resolved diagnostics for compression tuning and 
nonlinear correction. Most previous work, however, has focused on conventional accelerator 
beams, global compression settings, or dedicated linearization elements. For LWFA beams with an 
injection-induced broad energy spread, it remains less explored how a directly measured 
nonlinear LPS can be used to select the transmitted energy range and optimize the final bunch 
duration when first- and second-order longitudinal transport act simultaneously.

For femtosecond-scale electron beams, a terahertz transverse deflecting cavity (THz-TDC) 
provides a compact approach to LPS reconstruction with femtosecond-level temporal resolving 
power~\cite{yang2025sub,zhao2018terahertz,li2019terahertz,zhao2020femtosecond}.
In this work, a THz-TDC combined with a dipole magnet is employed to experimentally reconstruct 
the nonlinear LPS of LWFA electron bunches during double-bend-achromat (DBA)-based compression. 
The measurement directly resolves a characteristic C-shaped LPS induced by higher-order 
longitudinal transport. More importantly, rather than using the reconstructed LPS only to 
characterize the compression state, the measured nonlinear LPS provides an operational guide for 
aperture-defined energy-window control in the DBA. By varying both the central energy and the 
transmitted energy spread, different local-slope regions of the nonlinear LPS are selected, 
allowing the final bunch duration to be optimized under the combined influence of first- and 
second-order longitudinal transport. Space-charge-inclusive simulations reproduce the  
nonlinear LPS evolution and its energy-window-dependent bunch-duration behavior. The present 
work therefore advances LPS diagnostics from post-compression characterization to an 
experimentally accessible optimization strategy for compact LWFA beamlines, while also providing 
direct guidance for future higher-order LPS linearization.

The remainder of this paper is organized as follows. Section~\ref{sec:sim}
introduces the DBA-based compression principle and the role of energy-window
tuning, followed by numerical simulations of nonlinear LPS evolution and
bunch-length optimization. Section~\ref{sec:exp} describes the experimental
implementation of the THz-TDC diagnostic system, including THz pulse
characterization, temporal and energy resolution calibration, and LPS
reconstruction under different energy-window conditions. Section~\ref{sec:concl}
summarizes the main results and discusses their implications for optimizing
electron-bunch compression in LWFA-UED beamlines.

\section{\label{sec:sim}Numerical Simulation of DBA-Based Nonlinear LPS Evolution}

\subsection{\label{sec:dbaprinciple}Principle of DBA-Based Compression}

The DBA section provides magnetic bunch compression by introducing
energy-dependent longitudinal transport. In the dipoles, electrons with
different energies follow different trajectories and therefore acquire
different effective path lengths. This DBA-induced path-length difference
generates a longitudinal delay that can compensate the velocity-dependent
time-of-flight difference accumulated during beam propagation. The first-order
compression condition at the target position can be expressed
as~\cite{fang2022UED}
\begin{equation}
R_{56} = 2\rho\sin\alpha - 2\rho\alpha + \frac{L_z}{\gamma^2} \approx 0,
\label{eq:dba-condition}
\end{equation}
where \(\rho\), \(\alpha\), \(L_z\) and \(\gamma\) denote the dipole bending
radius, the bending angle, the effective propagation distance to the target
position, and the relativistic Lorentz factor of the reference particle,
respectively. The first two terms describe the geometrical path-length
contribution introduced by the DBA, while the last term represents the
velocity-dependent longitudinal dispersion accumulated during propagation.

\begin{figure}
    \centering
    \includegraphics[width=1\linewidth]{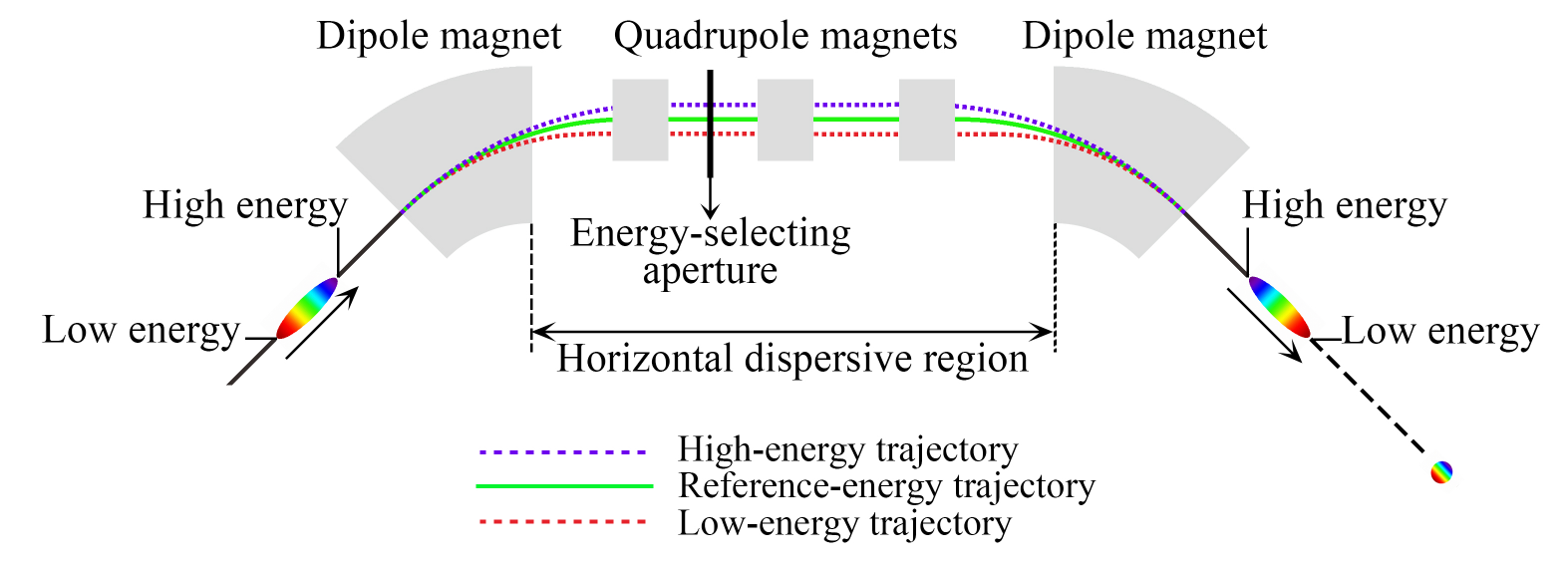}
    \caption{\label{fig:DBA}Schematic of DBA-based compression.}
\end{figure}

In the first dipole, electrons with different energies acquire
energy-dependent horizontal offsets in the dispersive region. An aperture
placed in this region defines the transmitted energy window and provides a
direct means of aperture-based phase-space control. Its transverse position
determines the energy-window center, while its opening controls the
transmitted energy spread. The intermediate quadrupole section provides
transverse focusing and optical matching to establish the horizontal
phase-space symmetry required for DBA transport, and the second dipole
recombines the trajectories at the DBA exit. During subsequent propagation,
higher-energy electrons travel slightly faster and catch up with lower-energy
electrons, thereby realizing first-order temporal compression at the target
position.

\subsection{\label{sec:simwindow}Simulation of Energy-Window-Dependent Nonlinear
LPS Evolution}

\begin{figure*}
    \centering
    \includegraphics[width=1\linewidth]{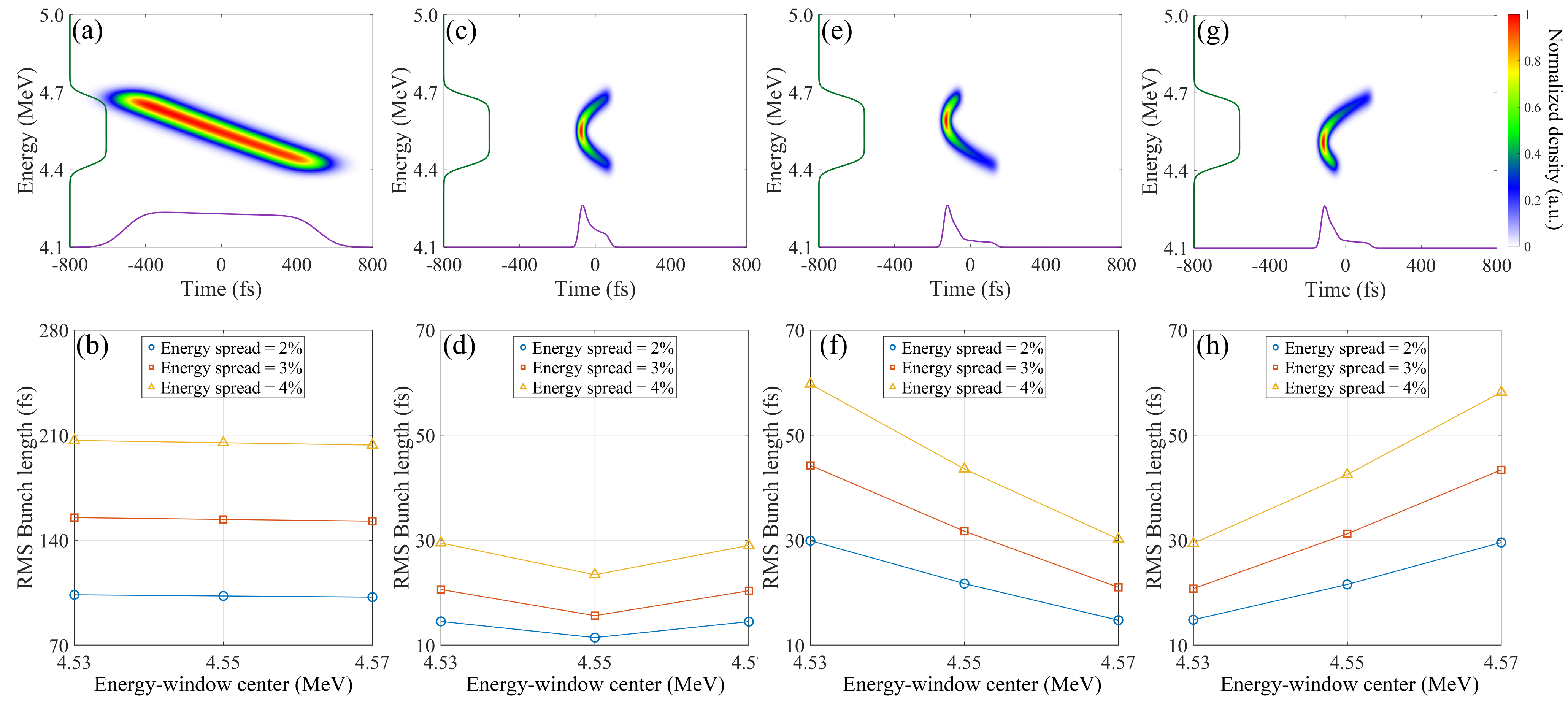}
    \caption{\label{fig:lps_sim}Numerical simulations of LPS evolution and
    energy-window-dependent bunch length. The simulated LPS distributions and
    corresponding RMS bunch lengths are shown at the target plane under
    different transport conditions: (a,b) without the DBA, and with the DBA at
    (c,d) the full-compressed condition, (e,f) the under-compressed condition, and (g,h) the over-compressed condition with appreciable second-order longitudinal transport. In all LPS plots, the left side of the time axis corresponds to
    the bunch head.}
\end{figure*}

To establish the phase-space basis for aperture-based compression optimization,
numerical simulations were performed to analyze the nonlinear LPS evolution
through the DBA section and the dependence of the final bunch length on the
energy-window condition. The simulation parameters were chosen to match the
experimentally relevant LWFA-UED beam conditions, while the initial \(6\%\)
full-width at half-maximum (FWHM) energy spread provides a sufficiently broad
phase-space range for applying the narrower energy windows used in the
following analysis. The simulations were performed with a bunch charge of
\(5~\mathrm{fC}\), a central energy of \(4.55~\mathrm{MeV}\) and an initial
root-mean-square (RMS) bunch length of \(10~\mathrm{fs}\), with space-charge
effects included. Figure~\ref{fig:lps_sim}(a) shows the simulated LPS after
drift-only propagation without the DBA. Although higher-order longitudinal
dispersion is also present in a drift section, the velocity-dependent
time-of-flight difference is dominated by the first-order term under this
condition, leading to pronounced temporal elongation while the nonlinear
curvature remains weak. With the DBA, the full-compression condition is considered first, where 
the first-order longitudinal dispersion \(R_{56}\) is reduced to near zero while second-order 
longitudinal transport remains appreciable, as shown in Fig.~\ref{fig:lps_sim}(c). For comparison, 
the under- and over-compression conditions are also considered, corresponding to positive and 
negative \(R_{56}\), respectively, as shown in Figs.~\ref{fig:lps_sim}(e) and 
\ref{fig:lps_sim}(g). Compared with the full compensated case, the nonzero first-order 
contributions in these two cases impose linear tilts of opposite signs on the nonlinear LPS and 
shift the low-slope region of the nonlinear LPS to different energy regions. 
For such a nonlinear LPS, the final temporal profile depends not only 
on the energy spread, but also on the position of the energy window within the LPS.

To quantify these effects, energy windows with different central energies and
energy spreads were applied to the LPS distributions. The central energies
were set to \(4.53\), \(4.55\), and \(4.57~\mathrm{MeV}\) to compare different
positions within the LPS, corresponding to the low-energy side, the
reference-energy region, and the high-energy side, respectively. The FWHM
energy spreads were set to \(2\%\), \(3\%\), and \(4\%\) to examine the
dependence of the RMS bunch length on the transmitted energy spread, as shown in 
Figs.~\ref{fig:lps_sim}(b), \ref{fig:lps_sim}(d),
\ref{fig:lps_sim}(f), and \ref{fig:lps_sim}(h).

The observed dependence can be interpreted using the local time--energy slope relation in 
Eq.~(\ref{eq:local-slope}).
As the energy-window position changes, the transmitted electrons occupy different local-slope 
regions of the nonlinear LPS; as the transmitted energy spread increases, the window covers a 
broader nonlinear phase-space range. 
Although the simulations include space-charge effects and the realistic beam distribution, these 
effects remain moderate for the $\mathrm{fC}$-level bunch charge considered here, so this 
transport picture provides a useful basis for identifying the dominant bunch-length dependence on 
the energy-window condition.
The trends observed in Figs.~\ref{fig:lps_sim}(b), \ref{fig:lps_sim}(d), \ref{fig:lps_sim}(f), 
and \ref{fig:lps_sim}(h) therefore reflect the
combined influence of the local time--energy slope and the width of the transmitted
energy window. The bunch length can be minimized when the transmitted energy window is aligned 
with a region of small local time--energy slope and the energy spread is sufficiently controlled. 
These results provide the phase-space
criterion used below to interpret the experimentally measured LPS distributions
and to optimize the aperture-defined energy-window condition.

\section{\label{sec:exp}THz-TDC-Based LPS Diagnostics}

\subsection{\label{sec:setup}Experimental Setup and THz Pulse Characterization}

\begin{figure*}
    \centering
    \includegraphics[width=1\linewidth]{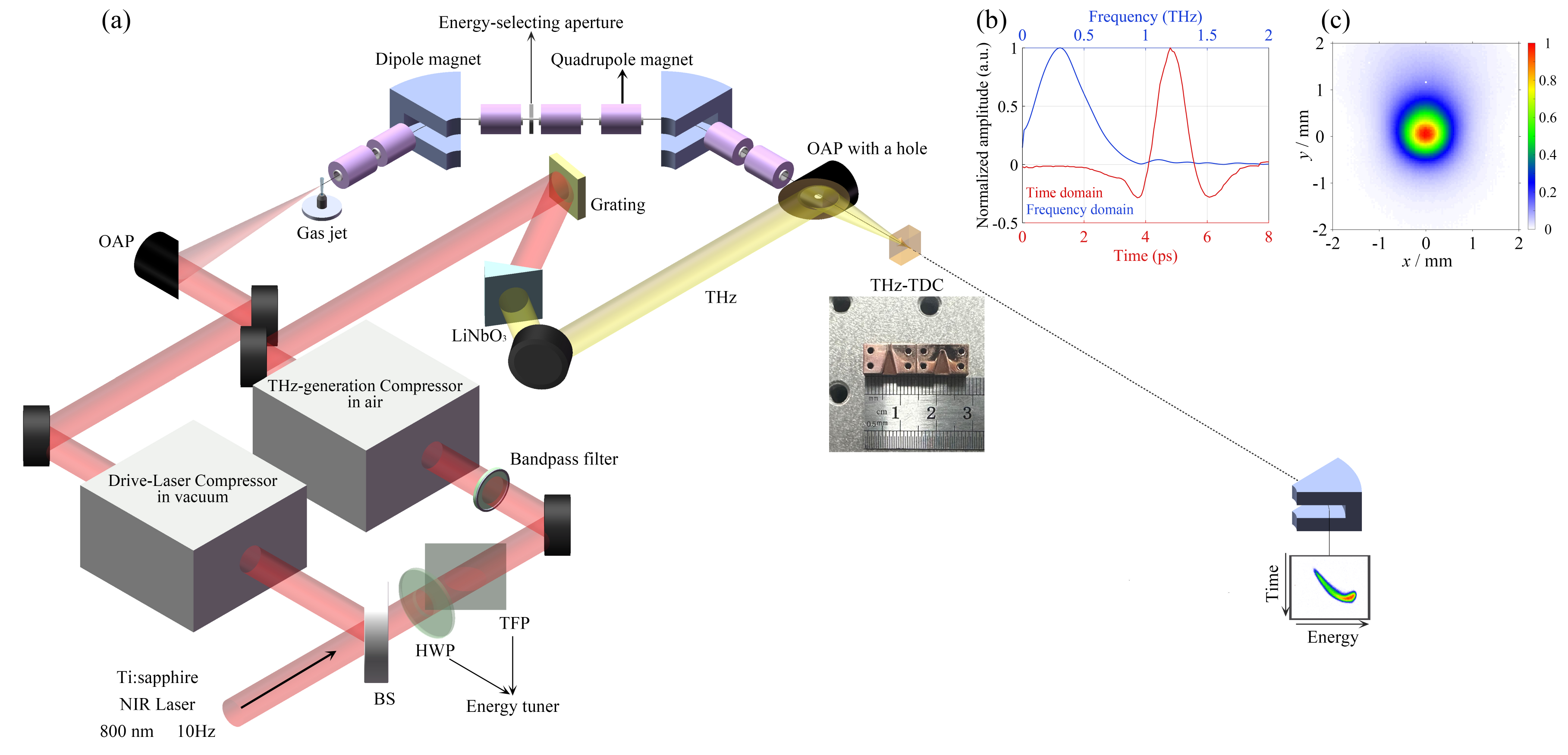}
    \caption{\label{fig:experiment}Experimental setup and THz pulse
    characterization. (a) Layout of the LWFA-UED beamline and THz-TDC
    diagnostics. (b) THz temporal waveform (red) and corresponding frequency
    spectrum (blue). (c) Measured THz focal spot at the electron--THz
    interaction point.}
\end{figure*}

The THz-TDC-based LPS diagnostic system was implemented on a compact LWFA-UED
beamline, as shown in Fig.~\ref{fig:experiment}(a). The system is driven by a
Ti:sapphire laser amplifier operating at a central wavelength of
\(800~\mathrm{nm}\) and a repetition rate of \(10~\mathrm{Hz}\). The output
pulse is split by a beam splitter (BS), with approximately \(40\%\) of the
energy directed to the LWFA branch. In this branch, the pulse is compressed to
\(24~\mathrm{fs}\) (FWHM) in a vacuum compressor and focused by an off-axis
parabolic (OAP) mirror into a nitrogen gas target to generate relativistic
electron bunches. The accelerated electrons are captured and focused by a
quadrupole doublet, pass through the DBA section for aperture-defined
energy-window control and longitudinal compression, and are subsequently
transported to the THz-TDC interaction point for LPS characterization.
The THz-TDC is based on a tapered rectangular waveguide fabricated 
in a split-block configuration, with the upper and lower halves machined 
separately and subsequently assembled.

The remaining \(\sim 60\%\) of the laser energy is directed to the THz
generation branch. In this branch, the pulse is compressed by an independent
compressor to a duration of approximately \(400~\mathrm{fs}\) (FWHM). To
optimize high-field THz generation via optical rectification in a lithium
niobate (\(\mathrm{LiNbO_3}\)) crystal, the THz-generation pulse undergoes
sequential energy and spectral conditioning, as shown in
Fig.~\ref{fig:experiment}(a). The pulse energy is first continuously adjusted
using a half-wave plate (HWP) and a thin-film polarizer (TFP), followed by a
bandpass filter to narrow the spectrum and optimize phase-matching conditions.

Under an incident laser energy of \(\sim 22~\mathrm{mJ}\), a single-pulse THz
energy of \(\sim 117~\upmu\mathrm{J}\) was achieved at the source, corresponding
to an optical rectification efficiency of approximately \(0.53\%\). After
transmission and focusing by an OAP with a hole, the THz pulse energy at the
interaction point was measured to be \(\sim 63~\upmu\mathrm{J}\). The THz temporal
waveform measured via electro-optic sampling (EOS) and the corresponding
frequency spectrum are shown in Fig.~\ref{fig:experiment}(b). The THz focal spot
at the electron--THz interaction point, with a beam waist radius of
approximately \(1~\mathrm{mm}\), is shown in Fig.~\ref{fig:experiment}(c).
Based on these parameters, the peak THz electric field is estimated
using~\cite{guiramand2022near}
\begin{equation}
E_{\mathrm{THz}}
=
\sqrt{
\frac{1}{c\varepsilon_0}
\frac{2W_{\mathrm{THz}}}{\tau A}
},
\label{eq:THz}
\end{equation}
where \(W_{\mathrm{THz}}\) is the THz pulse energy, \(\tau\) is the pulse
duration, and \(A\) is the focal spot area. The peak THz electric field is
thereby estimated to be approximately \(137~\mathrm{MV/m}\), providing
sufficient streaking strength for femtosecond-scale temporal diagnostics.

\subsection{\label{sec:resolution}Temporal and Energy Resolutions}

Before reconstructing the LPS, the temporal and energy resolutions were
calibrated separately. For the temporal axis, the THz-TDC maps the longitudinal
temporal coordinate of the electron bunch onto the vertical transverse
coordinate through THz streaking. The temporal resolving power is determined by
the unstreaked vertical beam size and the calibrated streaking strength. For
the energy axis, the THz field is turned off, and the downstream dipole magnet
maps the energy deviation onto the horizontal transverse coordinate. The energy
resolution is determined by the intrinsic horizontal beam size and the
horizontal dispersion at the detector plane.

\begin{figure}
    \centering
    \includegraphics[width=1\linewidth]{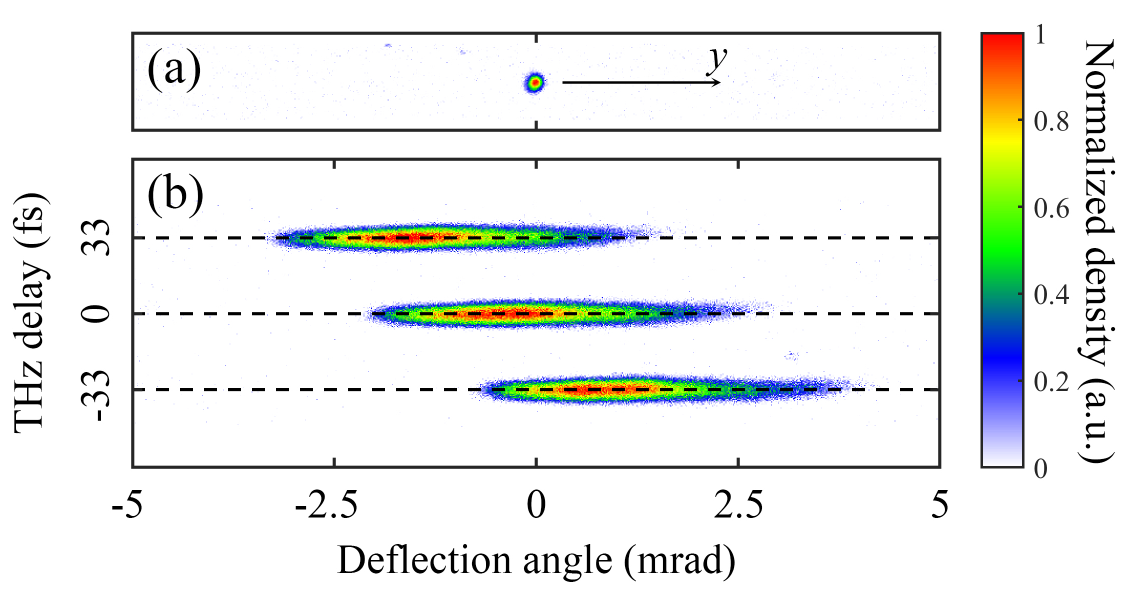}
    \caption{\label{fig:resolution}THz-TDC streaking calibration. (a)
    Unstreaked electron beam spot. (b) Streaked beam spot and calibration of
    the streaking strength.}
\end{figure}

The streaking strength was calibrated by scanning the relative delay between the
THz pulse and the electron bunch, as shown in Fig.~\ref{fig:resolution}. Near
the zero-crossing of the THz field, the beam centroid varies linearly with the
delay, yielding a calibrated streaking strength of
\(K \approx 36.2~\upmu\mathrm{rad}/\mathrm{fs}\). Given a drift distance of
\(L_d \approx 2.97~\mathrm{m}\) from the THz-TDC to the imaging screen, the
temporal resolution \(r_t\) is given
by~\cite{marx2018longitudinal,arpaia2020enhancing}
\begin{equation}
r_t = \frac{\sigma_{y_0}}{K L_d},
\label{eq:temporal_resolution}
\end{equation}
where \(\sigma_{y_0}\) is the unstreaked RMS vertical beam size. With
\(\sigma_{y_0} = 193 \pm 21~\upmu\mathrm{m}\), \(r_t\) is estimated to be
\(1.8~\mathrm{fs}\), with an uncertainty of approximately \(0.2~\mathrm{fs}\)
mainly arising from fluctuations in the vertical beam size.

The energy resolution \(r_E\) is expressed
as~\cite{dijkstal2024longitudinal}
\begin{equation}
r_E = E_0 \frac{\sigma_{x_0}}{D_x},
\label{eq:energy_resolution}
\end{equation}
where \(E_0\) is the average beam energy, \(D_x\) is the horizontal dispersion
at the detector plane, and \(\sigma_{x_0}\) is the unstreaked horizontal RMS
beam size. Using \(E_0 = 4.55~\mathrm{MeV}\),
\(D_x \approx 1.50~\mathrm{mm}/\%\), and
\(\sigma_{x_0} = 198 \pm 29~\upmu\mathrm{m}\), the energy resolution is estimated
to be \(6.0~\mathrm{keV}\), corresponding to a relative resolution of
\(0.13\%\). The uncertainty of this estimate, mainly arising from fluctuations
in the horizontal beam size, is approximately \(0.9~\mathrm{keV}\),
corresponding to a relative uncertainty of \(0.02\%\).

These calibrated resolutions are sufficient to resolve the femtosecond-scale
temporal structure and the MeV-scale energy correlation of the compressed LWFA
electron bunches. 

\subsection{\label{sec:lpsopt}LPS-Resolved Energy-Window Optimization}

\begin{figure*}
    \centering
    \includegraphics[width=0.9\linewidth]{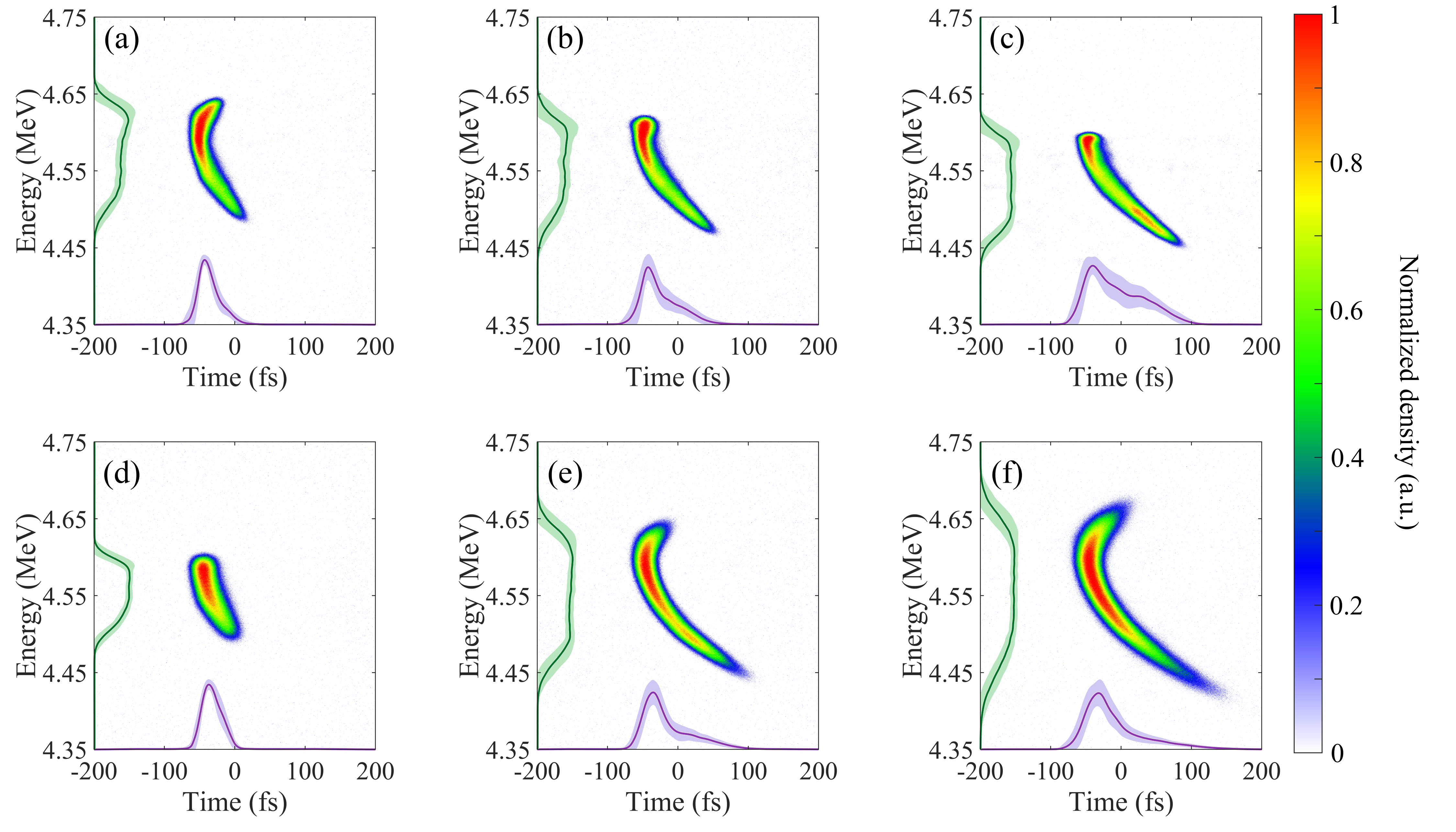}
    \caption{\label{fig:lps_measured}Measured LPS distributions under
    aperture-defined energy-window conditions. Panels (a)--(c) show
    measurements with comparable FWHM energy spreads of approximately
    \(2.8\%\)--\(2.9\%\) at central energies of \(4.574~\mathrm{MeV}\) (a),
    \(4.554~\mathrm{MeV}\) (b), and \(4.532~\mathrm{MeV}\) (c), respectively,
    illustrating the effect of energy-window position. Panels (d)--(f) show
    measurements near \(4.55~\mathrm{MeV}\) with FWHM energy spreads of
    \(2.0\%\) (d), \(3.8\%\) (e), and \(4.4\%\) (f), respectively, illustrating
    the effect of transmitted energy spread. The color maps show the
    reconstructed LPS density, with the green and purple marginal profiles
    denoting the energy and temporal projections, respectively. The shaded
    regions indicate the shot-to-shot standard deviation over 200 consecutive
    shots.}
\end{figure*}

\begin{table*}[t]
\caption{\label{tab:lps_conditions}
Energy-window conditions and measured RMS bunch lengths.
}
\begin{ruledtabular}
\begin{tabular}{cccc}
Case & Central energy & Energy spread (FWHM) & Bunch length (RMS) \\
\hline
a & \(4.574 \pm 0.007~\mathrm{MeV}\) & \(2.9 \pm 0.3\%\) & \(26 \pm 3~\mathrm{fs}\) \\
b & \(4.554 \pm 0.009~\mathrm{MeV}\) & \(2.8 \pm 0.5\%\) & \(35 \pm 5~\mathrm{fs}\) \\
c & \(4.532 \pm 0.008~\mathrm{MeV}\) & \(2.9 \pm 0.4\%\) & \(42 \pm 5~\mathrm{fs}\) \\
d & \(4.553 \pm 0.005~\mathrm{MeV}\) & \(2.0 \pm 0.2\%\) & \(27 \pm 3~\mathrm{fs}\) \\
e & \(4.552 \pm 0.010~\mathrm{MeV}\) & \(3.8 \pm 0.4\%\) & \(40 \pm 4~\mathrm{fs}\) \\
f & \(4.552 \pm 0.012~\mathrm{MeV}\) & \(4.4 \pm 0.6\%\) & \(44 \pm 6~\mathrm{fs}\) \\
\end{tabular}
\end{ruledtabular}
\end{table*}

To use the measured nonlinear LPS as a guide for aperture-based compression
optimization, LPS reconstruction was performed under different energy-window
conditions using the calibrated THz-TDC diagnostic system. The aperture
position and opening in the dispersive region controlled the central energy and
energy spread, respectively. Each measurement therefore represents the LPS of
electrons transmitted through a finite energy window. By scanning the aperture
position and opening, different portions of the nonlinear time--energy
correlation can be measured, and the full-energy-range LPS can in principle be
reconstructed by combining these window-resolved LPS measurements. The
following analysis therefore focuses on how the central energy and energy
spread of the transmitted window determine the temporal projection of the
bunch. Owing to the limited experimental space, the THz-TDC was positioned
slightly downstream of the designed full-compensation point, so the measurements correspond to an 
under-compressed state with appreciable second-order longitudinal transport. In this state, the 
low-slope region around the LPS inflection point is shifted toward the high-energy side. The 
measured LPS distributions under different energy-window conditions are shown in
Fig.~\ref{fig:lps_measured}. The corresponding energy-window conditions and
measured RMS bunch lengths are summarized in Table~\ref{tab:lps_conditions}.

The role of energy-window position was first examined by shifting the central
energy while maintaining a comparable energy spread of approximately
\(2.8\%\)--\(2.9\%\). Consistent with this picture, the measured C-shaped LPS shows that the 
energy window centered at $4.574\pm0.007$ MeV covers a region with a smaller local time--energy 
slope, whereas shifting the window toward $4.532\pm0.008$ MeV selects a region with a larger 
slope. Correspondingly, the RMS bunch length increases from $26\pm3$ fs to $42\pm5$ fs. This trend
shows that the aperture position controls the local time--energy slope covered by
the transmitted electrons.

The influence of transmitted energy spread was then examined with the central
energy kept near the reference-particle energy of \(4.55~\mathrm{MeV}\). As
summarized in Table~\ref{tab:lps_conditions}, the RMS bunch length shows an
overall increasing trend from \(27 \pm 3~\mathrm{fs}\) to
\(44 \pm 6~\mathrm{fs}\) as the FWHM energy spread increases from
\(2.0 \pm 0.2\%\) to \(4.4 \pm 0.6\%\). Although adjacent cases partially
overlap within their uncertainties, the trend indicates that increasing the
aperture opening includes a broader portion of the nonlinear LPS in the
temporal projection, leading to a longer bunch duration.

These measurements show that the reconstructed nonlinear LPS can serve as an
optimization map for DBA-compressed LWFA beams. The aperture position determines the 
local time--energy slope covered by the transmitted electrons, while the aperture opening 
determines the range of nonlinear phase space included in the temporal projection. Therefore, 
optimized compression requires both aligning the transmitted energy window with a
small-slope region of the nonlinear LPS and maintaining an appropriate energy
spread. This converts the LPS measurement
from a diagnostic of nonlinear transport into a practical guide for
aperture-based compression optimization in LWFA-UED beamlines.

\section{\label{sec:concl}Conclusion}

In summary, an integrated diagnostic system based on a THz-TDC and a dipole
magnet has been implemented for two-dimensional LPS reconstruction of LWFA
electron bunches during DBA-based compression. For electron bunches with an
average energy around \(4.55~\mathrm{MeV}\), the diagnostic system achieves a
temporal resolving power of \(1.8~\mathrm{fs}\) and an energy resolution of
\(6.0~\mathrm{keV}\), corresponding to a relative energy resolution of
\(0.13\%\). The system directly resolves a characteristic C-shaped LPS induced
by higher-order longitudinal transport, providing direct experimental access to
the nonlinear time--energy correlation that limits the final bunch duration. The measured LPS 
shows that optimized DBA compression requires simultaneous matching of the beam central energy, 
the first-order longitudinal dispersion compensation, and the energy spread. The first-order 
dispersion \(R_{56}\) determines the main compression condition and the local time--energy slope 
near the beam center, while the second-order term \(T_{566}\) introduces nonlinear curvature that 
can broaden the temporal projection even when the first-order elongation is nearly compensated. 
Therefore, short-bunch operation requires the beam to be centered in a low-slope region of the 
nonlinear LPS, with the energy spread sufficiently controlled to avoid excessive sampling of the 
curved phase-space region. 

Beyond identifying the optimal operating condition in the present 
experiment, the LPS measurement provides direct guidance for DBA beamline design. It can be used 
to evaluate whether the designed first-order dispersion compensation matches the actual beam 
central energy, to determine the acceptable energy-spread range, and to assess the need for 
higher-order longitudinal correction. In particular, the measured C-shaped LPS provides an 
experimental basis for future \(T_{566}\) compensation or linearization, for example by 
introducing sextupole magnets in the dispersive section~\cite{england2005sextupole}. 
Such LPS-guided matching of first- and second-order longitudinal transport could further improve 
bunch compression and support the development of LWFA-UED toward sub-\(10~\mathrm{fs}\) temporal 
resolution.

\begin{acknowledgments}
The authors would like to acknowledge Dr. Xiaonan Ning and Dr. Jiucheng Chen 
for their valuable assistance in optimizing the laser system; Dr. Wentao Yu 
and Kai Peng for their assistance with the setup  of the THz system; and 
Dr. Yuemei Tan and Peng Lv for helpful discussions on the THz-TDC. 
The authors also gratefully acknowledge financial support from the Strategic Priority Research
Program of the Chinese Academy of Sciences (Grant No.~XDB0530000), Beijing
Natural Science Foundation (Grant No.~L2602047), the National Natural Science
Foundation of China (Grant No.~12405169) and the Discipline Construction
Foundation of the ``Double World-Class Project''. The numerical simulations
were carried out using the computational resources provided by the High
Performance Computing Center of Tsinghua University.
\end{acknowledgments}

\bibliography{aipsamp}

\end{document}